\documentclass[12pt]{article}

\usepackage{epsfig,rotating}
\usepackage{cite}

\newcommand{\beq}{\begin{equation}}
\newcommand{\eeq}{\end{equation}  }

\newcommand{\bec}{\begin{center}}
\newcommand{\eec}{\end{center}}

\def\d{\delta}

\newcommand{\W}{\mbox{w}}
\newcommand{\WW}{\mbox{ww}}

\begin{document}
\clearpage
\pagestyle{empty}
\setcounter{footnote}{0}\setcounter{page}{0}%
\thispagestyle{empty}\pagestyle{plain}\pagenumbering{arabic}%

\vspace{2.0cm}

\begin{center}

\vskip 0.8in plus 2in

\hfill ANL-HEP-CP-00-03 

\hfill January 10, 2000 

\vspace{3.0cm}

%
%

{\Large\bf Bose-Einstein Correlations in \boldmath{$e^+e^-\to W^+W^-$}  \\ 
at a Linear Collider\\[-1cm]}

\vspace{3.0cm}

{\large  S.~V.~Chekanov$^a$, A.~De Roeck$^b$ and E.A.~De~Wolf$^c$}

{\begin{itemize}
\itemsep=-1mm
 
\normalsize
\item[$^a$]
 
\small
Argonne National Laboratory,
9700 S. Cass Avenue,
Argonne, IL 60439
USA

\normalsize
\item[$^b$]
\small Deutsches Electron-Synchrotron DESY, 
Notkestrasse 85, D-22603, Hamburg, FRG\\
and CERN, 1211 Geneva 23, Switzerland

\normalsize
\item[$^c$]
 
\small
Department of Physics, Universitaire  Instelling Antwerp,
B-2610 Wilrijk, Belgium

\end{itemize}
}
 
\normalsize
\vspace{1.0cm}

Proceedings  of the Workshop "Physics  
Studies for a Future Linear Collider", \\ 
QCD Working Group, 2000, DESY 123F 

\vspace{1.0cm}

\begin{abstract}
\noindent We show that the most popular  method to simulate 
Bose-Einstein (BE) interference effects predicts negligible 
correlations between identical pions originating from the hadronic decay
of different  $W$'s produced in  
$e^+e^- \to W^+W^- \to 4 \;\hbox{\rm jets}$ at  typical
linear collider energies. 
\end{abstract}

\end{center}

\setcounter{page}{1}

\noindent It is known that the Bose-Einstein (BE) effect can produce a
systematic uncertainty on the $W$ mass measurement in   
the process $e^+e^- \to W^+W^- \to 4 \;\hbox{\rm jets}$. 
For the LEP2 experiments, this uncertainty can be 
serious~\cite{sjo} if it is
as large as  expected from the 
standard BE simulation included in the PYTHIA/JETSET 
Monte Carlo model~\cite{jet}.  
Considering that the TESLA linear collider (LC) will make it possible to study
$e^+e^-$ annihilation 
with  much larger statistics  and at  much higher centre-of-mass-energy,
it is important to understand how strongly  
the $W$-mass measurement at a LC could be
affected by the BE effect.

In this study we use the most popular and simplest
BE simulation technique  based on  the LUBOEI algorithm, 
as included in  PYTHIA/JETSET  model.
It is known that this algorithm  has a strong effect on the
$W$-mass reconstruction and the systematic uncertainties
on the $W$ mass can be as large as 50 MeV  at LEP2 energies\cite{todo}.   
Other methods of BE simulation usually show  much smaller
uncertainties  at LEP2 energies (see~\cite{rev} for
 recent reviews).  

In this paper we investigate the BE correlations 
at TESLA energies   
using  the tools proposed in~\cite{div1} and~\cite{sub1},
without attempts to calculate shifts of the $W$ mass
for a particular reconstruction method. 
By comparing the behaviour of the correlations with those 
observed at LEP2 energies, quantitative information can be obtained
on possible effects on $W$ mass measurements.
 
To analyse the BE correlations, we use the following 
two methods: 

1) Division method~\cite{div1}. Here the correlations are
measured using the ratio:
\beq
R^*=\frac{\rho^{\WW}(\pm,\pm) - 2\,\rho^{\W}(\pm,\pm)}
{\rho^{\WW }(+,-) - 2\,\rho^{\W}(+,-)}, 
\label{del1}
\eeq
where $\rho^{\WW}$ and $\rho^{\W}$ are the two-particle densities
for 4 and 2-jet hadronic $W$ decays, respectively,  and $(\pm, \pm )$ and $(+, -)$
denote like-charge and unlike-charge particle combinations.
With this  definition, $R^*$ is unity if there is  no cross-talk between  two $W$'s.
Thus $R^*$ can be used as an indicator of BE  
interference between hadrons from different $W$ bosons. 
We note that this quantity resembles the standard BE correlation function
when unlike-charged particles are used as a reference,
but has, in fact, little to do with it~\cite{sub1}.

2) Subtraction method~\cite{sub1}:
\beq
\d\rho = 
\rho^{\WW}(\pm,\pm) - 2\,\rho^{\W}(\pm,\pm) -
\rho^{\WW}(+,-)+2\, \rho^{\W}(+,-).  
\label{dif}
\eeq
In the absence of cross-talk between  $W$'s, one has 
$\d\rho =0$. In the following we  use $1+\d\rho$,
rather than~(\ref{dif}) in analogy with $R^*$.    

The two methods differ mainly in the fact that,
for $\d\rho$   the so-called
mixing terms (terms determined by the product of single-particle
densities) cancel, whereas such mixing terms
are still present in the method~1, and largely determine the 
behaviour of $R^*$ in the case of cross-talk between $W$ bosons~\cite{sub1}.

Results for both methods, using the LUBOEI algorithm,  
have been presented for LEP2 energies in~\cite{sub1}. 
About 20K WW events were generated at $\sqrt{s}=190$ GeV. 
A clear enhancement of $R^*$ and $\d\rho$ was observed when the
squared 4-momentum  difference $Q_{12}\equiv \sqrt{-(p_1 - p_2)^2}$ between
two like-sign particles decreases (see Fig.~8 and 9 of~\cite{sub1}).

Here we present results of a similar study at $\sqrt{s}= 180$~GeV
and $\sqrt{s}=500$~GeV,
for a LC.
The LC is expected to operate at $\sqrt{s} = 500$ GeV most of the time,
collecting an integrated luminosity of 200 fb$^{-1}$ per year. The 
TESLA proposal for the LC will include an option to run at smaller
beam energies, down to $\sqrt{s} = 90$ GeV, but with reduced 
luminosity. If the physics motivation is 
sufficiently strong, TESLA can also run at energies slightly above
the $WW$ threshold, e.g. at $\sqrt{s} = $ 180 or 200 GeV.
Event samples which can reasonably expected to be collected at this energy,
amount to 10-20 fb$^{-1}$, i.e. about 20 times larger than 
the data samples collected at LEP2.

\begin{figure}[htbp]
\vspace{-1.5cm}
\begin{center}
\mbox{\epsfig{file=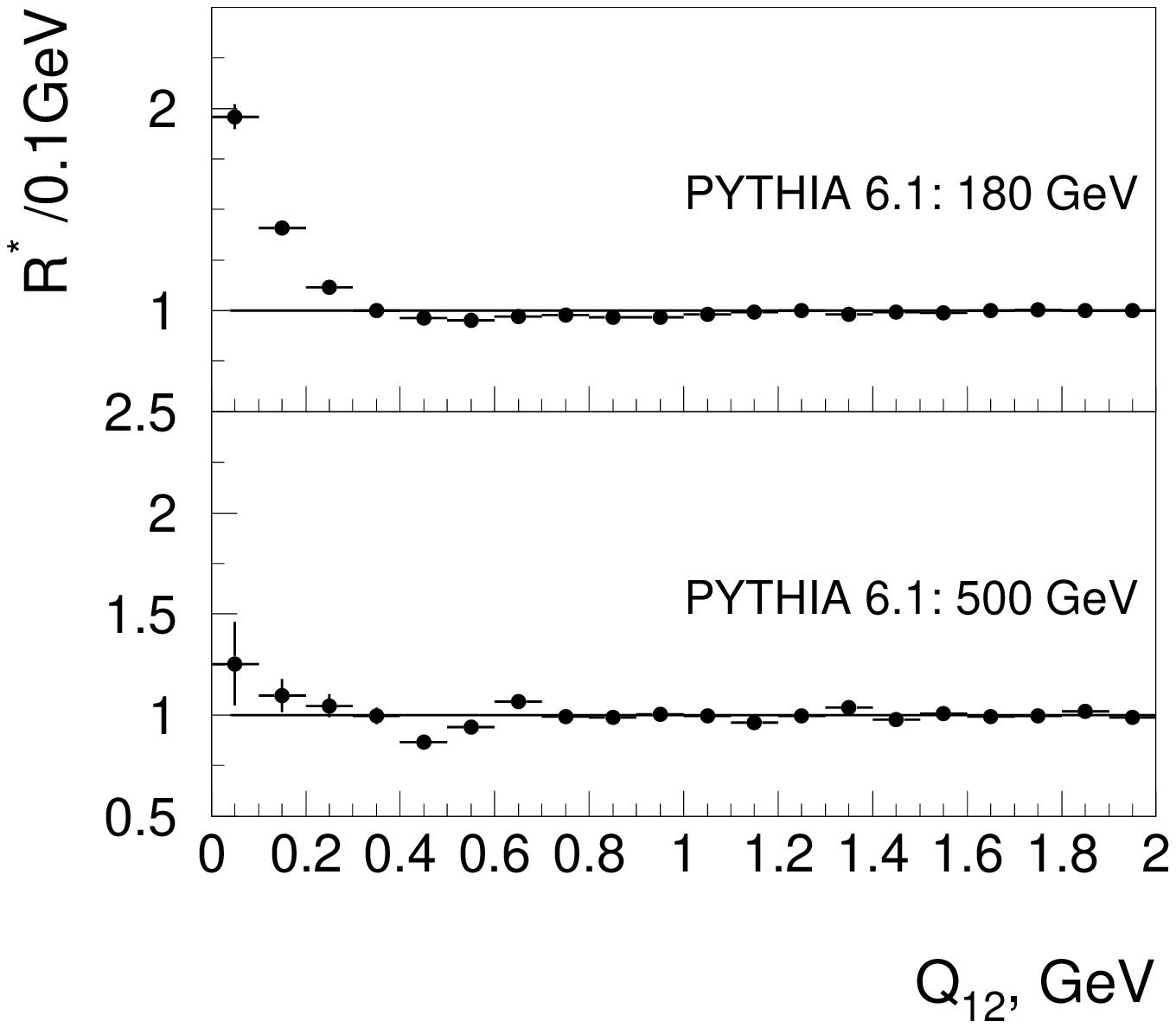, width=0.6\linewidth}}
\end{center}
\caption{
$R^*$ for PYTHIA Monte Carlo model  without and with  
BE correlations at $\sqrt{s}=180,500$ GeV. The solid line
corresponds to the model without BE effect.}
\label{8j}
 
\begin{center}
\mbox{\epsfig{file=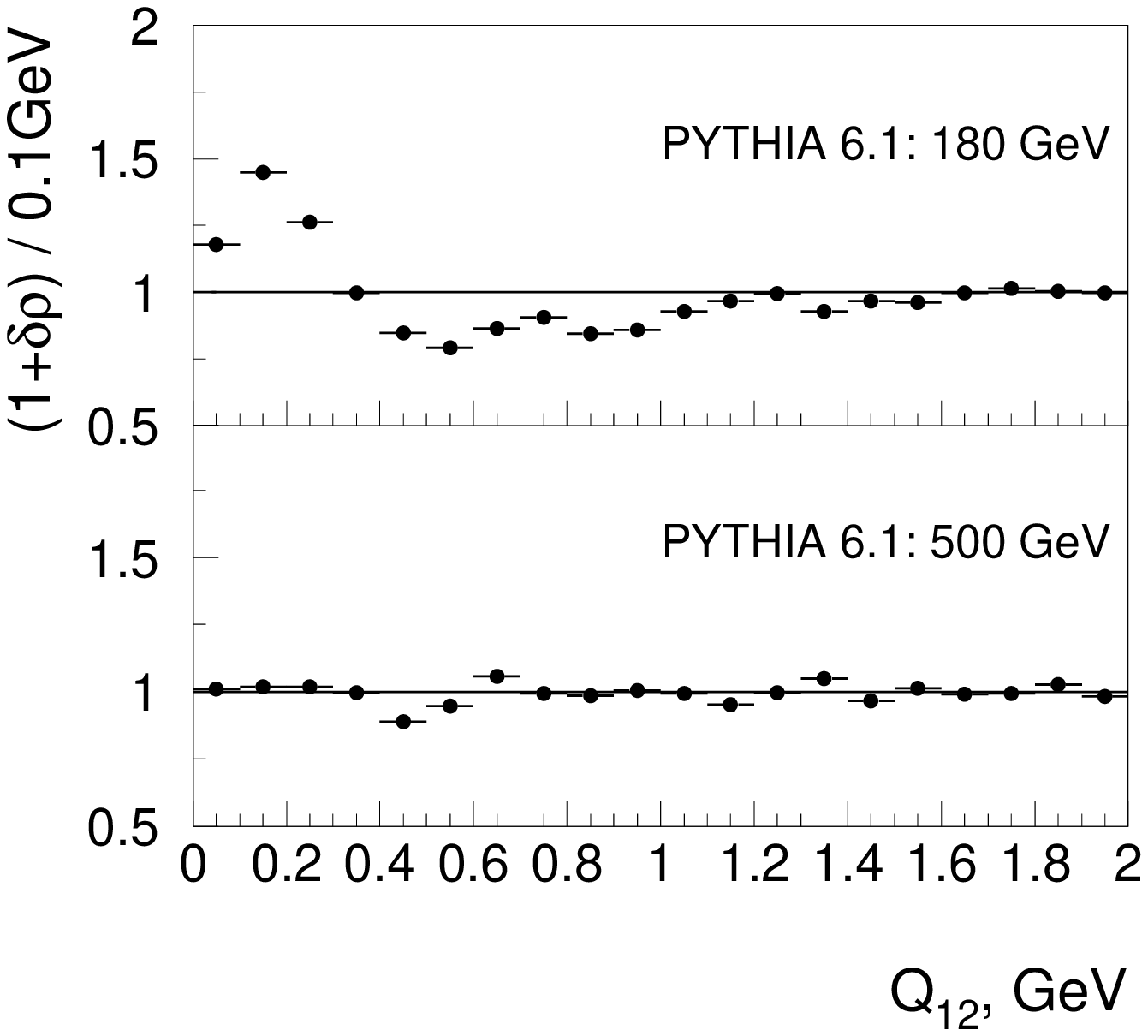, width=0.6\linewidth}}
\end{center}
\caption{
$1+\delta\rho$ for PYTHIA  Monte Carlo model  without
and with BE correlations at $\sqrt{s}=180,500$ GeV. The solid line
corresponds to the model without BE effect.}
\label{9j}
\end{figure}

The results for $R^*$ and  
$1 + \d\rho$ are shown in Fig.~\ref{8j} and Fig.~\ref{9j}. The solid lines in the
figures   correspond 
to  the case of no BE correlations. 
At each energy 200K $WW$ events were generated,  
corresponding  to a luminosity of 15 fb$^{-1}$ and 30 fb$^{-1}$ 
at  $\sqrt{s}=180$ GeV and $\sqrt{s}=500$ GeV, respectively.

The figures show no increase  of   $R^*$ and $1 + \d\rho$ 
for small $Q_{12}$ at  $\sqrt{s}=500$ GeV, whereas 
BE correlations are clearly seen  at 180 GeV.
At an intermediate energy, $\sqrt{s}=300$ GeV, both correlation
quantities have also been studied (not shown). 
The BE effect was found to be
much smaller than  that at $\sqrt{s}=180$ GeV.
Therefore, in order to detect this  effect at $\sqrt{s}=300$ GeV
much  higher statistics than at LEP is needed.  

The statistics used for the  present paper  are  
ten times larger than that used for LEP2 studies~\cite{sub1}.
The figures illustrate that, with the 
much higher statistics expected at LC
more details in the
behaviour of  BE correlations can be observed, especially for 
$1 + \d\rho$. This correlation
function is slightly below unity for $Q_{12}\sim 0.3-1.2$ GeV 
and has a rather complex structure. This is much less evident   
for $R^*$, where  the details  of the genuine correlations
are hidden due to the presence of large mixing terms 
in the definition~(\ref{del1}).   

Our results demonstrate that BE correlations
between hadrons from different $W$'s 
can be studied  at energies close to the $WW$ threshold.
At much higher  energies the effect is negligible. This is
related to the
large phase space available for  secondary particles
from hadronic $W$ decays at high energy, and the small  probability  for 
two identical secondary particles from different $W$'s
to be emitted close in $Q_{12}$ -phase space\footnote{A similar 
observation was made in~\cite{schieck}}.

In conclusion, studies at $\sqrt{s}=500$ GeV do not indicate the presence of
 Bose-Einstein correlations
between identical hadrons from decays of  different $W$ bosons for
realistic statistics attainable at a LC. The systematic uncertainty on a measurement
of the $W$ mass caused by these correlations is therefore expected to be  negligible.
This result is based on the most popular
method to simulate BE correlations (LUBOEI in PYTHIA/JETSET)
and two different experimental techniques.  
The LUBOEI algorithm 
is known to produce the largest 
effect on the $W$ mass reconstruction at LEP2 energies.
It is therefore reasonable to expect that the use of other 
available models, known to lead to smaller
uncertainties on the
reconstruction of $W$ mass, can only strengthen the above conclusion.
However when the LC will be operated at an energy close to but above the 
$WW$ threshold, collecting event samples of the order of 10-20 fb$^{-1}$,
BE effects between different $W$'s can be studied with uncanny precision. 

{}

\newpage

\end{document}